\documentclass{eptcs}

\usepackage{iftex}
\usepackage{graphicx}
\usepackage{booktabs}
\usepackage{amsmath}

\newtheorem{example}{Example}[section]

\newtheorem{theorem}{Theorem}[section]

\newcommand{\constantSet}{\mathcal{C}}

\newcommand{\predicateSet}{\mathcal{P}}
\newcommand{\variableSet}{\mathcal{V}}

\newcommand{\groundLiteralSet}{\mathcal{G}}
\newcommand{\interpretationSet}{\mathcal{I}}
\newcommand{\interpretation}{I}
\newcommand{\semiLattice}{\mathcal{L}}
\newcommand{\delay}{\Delta t}

\newcommand{\program}{\Pi}
\newcommand{\fixpointOperator}{\Gamma}
\newcommand{\groundedLiteral}{g}

\newcommand{\nonGroundedformula}{f}

\def\domainSet{\mathcal{D}}

\def\atPred{\texttt{at}}
\def\connPred{\texttt{conn}}

\def\after{\textsf{AFTER}}
\def\before{\textsf{BEFORE}}

\newcommand{\conv}{\ast}
\newcommand{\prevalue}{value}

\def\anomPred{\texttt{anomalyType}}

\ifpdf
  \usepackage{underscore}         
  \usepackage[T1]{fontenc}        
\else
  \usepackage{breakurl}           
\fi

\title{Geospatial Trajectory Generation via Efficient Abduction: Deployment for Independent Testing}
\author{Divyagna Bavikadi \quad Dyuman Aditya \quad Devendra Parkar \quad Paulo Shakarian
\institute{
Arizona State University\\
Arizona, USA
}
\email{\quad dbavikad@asu.edu \quad daditya1@asu.edu \quad dparkar1@asu.edu \quad 	pshak02@asu.edu }
\and
Graham Mueller 
\institute{ Leidos \\
Virginia, USA
}
\email{William.G.Mueller@leidos.com }
\and
Chad Parvis
\institute{ EpochGeo\\
Washington,D.C, USA}
\email{ cp@epochgeo.com  }
\and
Gerardo I.\ Simari 
\institute{Universidad Nacional del Sur (UNS) \\
Bahía Blanca, Argentina}
\institute{CONICET\\
Bahía Blanca, Argentina}
\email{gisimari@gmail.com}
}

\def\titlerunning{Geospatial Trajectory Generation}
\def\authorrunning{D. Bavikadi et al.}

\begin{document}
\maketitle

\begin{abstract}
The ability to generate artificial human movement patterns while meeting location and time constraints is an important problem in the security community, particularly as it enables the study of the analog problem of detecting such patterns while maintaining privacy.  We frame this problem as an instance of abduction guided by a novel parsimony function represented as an aggregate truth value over an annotated logic program.  This approach has the added benefit of affording explainability to an analyst user.  By showing that any subset of such a program can provide a lower bound on this parsimony requirement, we are able to abduce movement trajectories efficiently through an informed (i.e., A*) search.  We describe how our implementation was enhanced with the application of multiple techniques in order to be scaled and integrated with a cloud-based software stack that included bottom-up rule learning, geolocated knowledge graph retrieval/management, and interfaces with government systems for independently conducted government-run tests for which we provide results.  We also report on our own experiments showing that we not only provide exact results but also scale to very large scenarios and provide realistic agent trajectories that can go undetected by machine learning anomaly detectors.
\end{abstract}

\vspace{-1em}
\section{Introduction}
The ability to generate artificial human movement patterns while meeting location and time constraints is an important problem in the security community, particularly as it enables the study of the analog problem of detecting such patterns without the need for data from actual humans.  This work is part of a larger effort to establish models of normal human movement at a fine-grain level\footnote{The IARPA HAYSTAC program, \url{https://www.iarpa.gov/research-programs/haystac}} and operationalize those models and techniques in a system deployed to a government environment for evaluation.  This contrasts with current techniques used to model populations that operate at a more coarse-grain level (country, county-level than building-level) as seen in previous work in specific areas such as population migration~\cite{alis2021generalized} or disease spread \cite{doi:10.1073/pnas.2313171120}.

In this work, we focus on the generation of human movement patterns based on historical data.  We frame this problem as an instance of abduction~\cite{peng-reggia90} in a geographic setting~\cite{shak11} guided by a novel parsimony function represented as an aggregate truth value over an annotated logic program~\cite{ks92}.  This approach has the added benefit of affording explainability to an analyst user.  By showing that any subset of such a program can provide a lower bound on this parsimony requirement, we are able to abduce movement trajectories efficiently through an informed (i.e., A*) search.  This fundamental result enables the practical implementation and deployment of software that is independently evaluated by a government test and evaluation (T\&E) team. Specifically, in this paper, we make the following contributions.
\begin{enumerate}
    \item We leverage the modularity of a logic program to guide informed search.  Specifically, we provide a general result where we prove that a parsimony function consisting of an aggregate truth value of a logic program is bounded by such an aggregate of a subset of the program.  In the case of our application, when rules correspond to one hop in an underlying graphical structure (in our case, a road network) we can provide an efficient graph-based heuristic based on above general result.
    We also provide empirical findings illustrating that the lower bound of the parsimony function for a subset logic program as well as show that an A* implementation provides improvements not available without a heuristic.
    \item We build on the above strategy to provide scale through ad-hoc creation of the heuristic function that proceeds with the search.  We provide empirical results demonstrating the scalability of the approach.
    \item We show that our approach generates movement trajectories that are robust to machine learning (ML) models designed specifically to detect anomalous movements.  We show the results of an internal test against such an ML model where we compare the output of the model with that of the training data and find in the vast majority of cases trajectories produced by our approach have anomaly ratings at or below the training data.  We also illustrate how our approach allows for explainability of the anomalous portions of a trajectory by leveraging the deduction results of the logic program.
    \item We describe how our approach can be integrated with rule mining, graph databases, and Amazon Web Services (AWS) cloud infrastructure in a system deployed for government testing. Further, we provide results of the government test where our approach was evaluated in four different environments against nine different machine learning models, all designed to find our trajectories.  In the majority of cases, the ML models find our trajectories with a probability of detection (PD) of less than $0.40$, which is the standard established by the government.    
    \end{enumerate}
The remainder of the paper is organized as follows.  In Section~\ref{sec:backgroundTgtApp} we describe the application domain, and Section~\ref{sec:prelims} provides a review of requisite logical preliminaries previously introduced in \cite{ks92,ssTAI22,aditya2023pyreason}. This is followed by Section~\ref{sec:theory}, where we describe framing abduction for this domain by bounding the parsimony function to warrant informed search. The deployment is shown in Section~\ref{sec:software}. We report experimental results in Section~\ref{sec:internaleval} and discuss the findings of our internal evaluations where we show (in practice) tractable yet exact computation of parsimonious explanations, that movement trajectories provided are not susceptible to detection by machine learning-based anomaly detectors, and further scalability to larger graphs.
In Section~\ref{sec:externaleval} we describe an independent evaluation with the probability of detection, Section~\ref{sec:relwork} presents related work, and we conclude in Section~\ref{sec:conclusion} by outlining future research. 

\vspace{-1em}
\section{Motivating Application}
\label{sec:backgroundTgtApp}

\begin{figure*}[t]
\centering
        \includegraphics[height=2in]{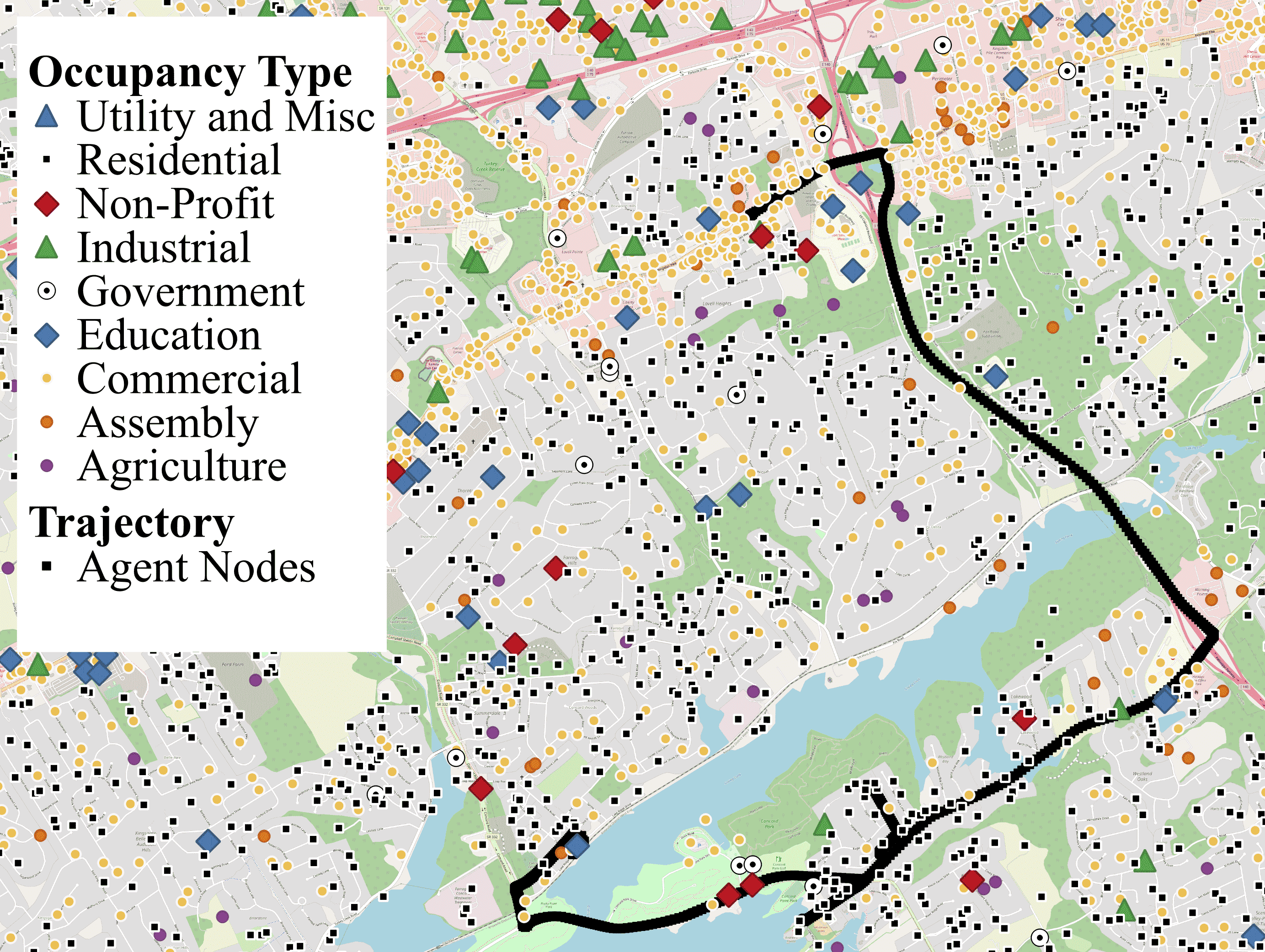}
        \includegraphics[width=2.98in, height=2in]{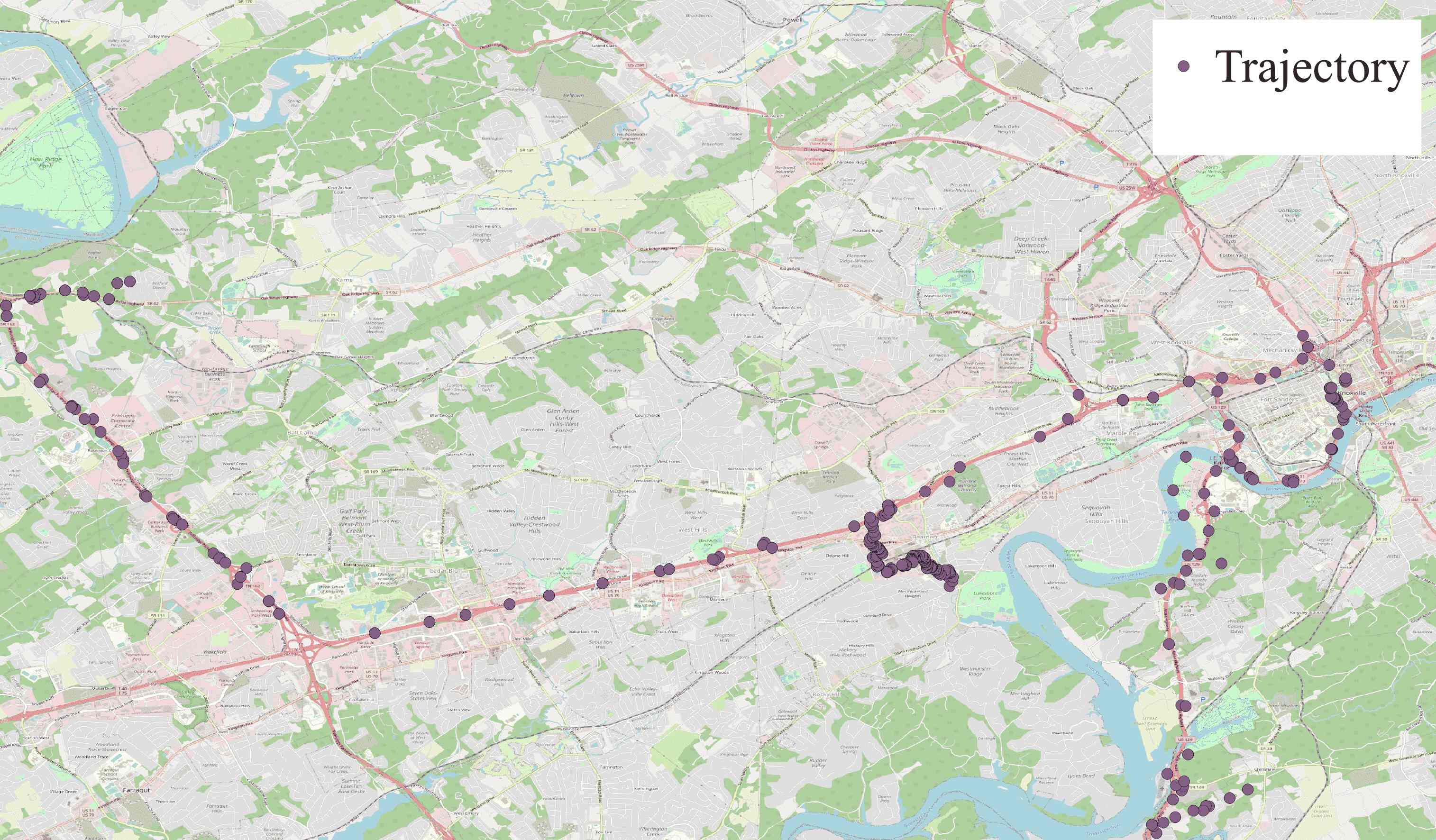}
       
    \caption{Left: The graph represents the city of Knoxville with landmarks, and the line plot denotes part of an agent's sub-trajectory. Right: 
    Generated trajectory.}
     \label{fig:data}
\end{figure*}

In the aftermath of an unexpected event, such as a natural disaster, war, or large-scale industrial accident, human movement patterns can change significantly.  As a result, IARPA (Intelligence Advanced Research Projects Activity) has identified problems relating to the characterization and generation of normal human movement patterns as a key problem of study in the HAYSTAC program\footnote{https://www.iarpa.gov/newsroom/article/finding-a-needle-with-haystac}.  In this problem, a given geospatial area is modeled as a graph where locations (we shall use $\domainSet_{loc}$ to denote the domain of all possible locations) represent the vertices.  A set of agents, e.g., $007, 008,.. \in \domainSet_{agent}$, traverses through the network by using various modes of transportation, such as  $personal\_vehicle \in \domainSet_{movtype}$.  Throughout the paper, we will use annotated logic~\cite{ks92} to specify various aspects of the environment by assigning predicates and then use temporal extensions~\cite{aditya2023pyreason}; temporal rules to specify the normalcy or abnormalcy of an agent's movement throughout the geospatial area.  The rules will be learned from historical data, hence the abduction problem will consist of producing an agent trajectory (an explanation) between a start and end locations at certain times (observations) such that the abnormalcy of the trajectory is minimized (parsimony requirement).  An example of a movement trajectory (before interpolation, discussed in Section~\ref{sec:externaleval}) produced by our approach is shown in Figure~\ref{fig:data}(right) for the same agent whose training trajectory is shown in Figure~\ref{fig:data}(left).  However, despite the specific application and ensuing deployment, we describe, our key insight is more general -- if the parsimony function is specified as an aggregate over truth values resulting from the deductive process of a logic program, we can leverage the logic program to correctly improve efficiency (discussed in Section~\ref{sec:theory}).  This insight is what enables us to ultimately solve the problem at scale.

\vspace{-1em}
\section{Technical Preliminaries}
\label{sec:prelims}

\noindent\textbf{Syntax of Annotated Logic with Temporal Extensions.} We use annotated logic syntax~\cite{ks92} with a lower-lattice based semantics~\cite{ssTAI22} and temporal syntactic and semantic extensions~\cite{aditya2023pyreason}.  
We assume a set of constant, variable, and predicate symbols (resp., $\constantSet$, $\variableSet, \predicateSet$) where the set of constants is divided into domains (i.e., $\domainSet_i \subset \constantSet$).  Atoms are formed with terms (constants or variables) and predicates.  Literals include atoms and their negation, with $\groundLiteralSet$ being the set of all (ground) literals formed with no variables.  Atoms and literals can be annotated with elements (intervals in $[0,1]$) of a lower semi-lattice structure  $\semiLattice$ (not necessarily complete) with ordering $\sqsubseteq$.  Here, to generalize fuzzy logic in the spirit of~\cite{NilssonProbabilisticLogic}, we define the bottom element $\bot$ as $[0,1]$ and a set of top elements $\{[x,y] \; | \; x=y\}$.  If $\mathbf{a}$ is an atom and $\mathbf{\mu}$ is an annotation that denotes truth probability, then $\mathbf{a:\mu}$ is an \emph{annotated atom}. We can create annotated negations (of atoms) in the same way, and together with annotated atoms this gives us the set of \emph{annotated literals}.  We also find it useful to extend the logic to make statements about time, and we do this in two different ways.  First, following the convention of~\cite{shakarian2012annotated} we have \textit{temporally annotated facts (TAFs)}.  Given time point $t$ and annotated literal $f$, $f_t$ is a TAF that is true at $t$.  The second extension is similar to that of Allen's temporal logic~\cite{ATL} in which we have modal operators $\after(f,f')$ and $\before(f,f')$, where $f,f'$ are annotated literals, and the intuition is that $f$ occurs after (resp., before) $f'$.  Annotated literals and constructs formed with $\after$ and $\before$ are called annotated formulas.
Note that TAFs are considered separately from these formulas when we describe the semantics.

\begin{example}[Language and Syntax]
\label{ex:syn}
\textit{We define specific domains as follows: $\constantSet$ consists of disjoint sets $\domainSet_{agent}, \domainSet_{loc}, \domainSet_{movtype}$. $\domainSet_{agent}$ are constants associated with agents in the environment whose behavior we wish to model as mentioned in Section~\ref{sec:backgroundTgtApp}. We will have various predicates such as the binary predicate $\connPred$ that takes elements of $\domainSet_{loc}$ as arguments representing the connection of two locations.  The truth values associated with ground atoms created with this predicate specify a road network (as depicted in Figure~\ref{fig:data}).  We will also have a binary predicate, $\atPred$ specifying that an agent is at a certain location, e.g., for agent $agent$ and location $loc \in \domainSet_{loc}$, $\atPred(agent,loc)$ is a ground atom. We also have various unary predicates such as $\texttt{prim-Banks}$, $\texttt{occ-Education}$, $\texttt{occ-Residential}, ...$ (where $prim$ encodes the primary use of the location like the building is primarily used as a $Bank$, and $occ$ implies the occupancy type of the building like for $Residential$ purpose) that take elements of $\domainSet_{loc}$ as an argument.  Finally, predicates of the form $\anomPred(agent)$ are concluded from rules and are considered true with confidence indicated by its annotation if the agent is conducting abnormal behavior.}
\end{example}

\medskip
\noindent\textbf{Semantics and Satisfaction.}  Following previously introduced temporal extensions to annotated logic (i.e., \cite{aditya2023pyreason,shakarian2011annotated}), we assume a finite series of time points that we wish to reason about in an associated semantic structure of an interpretation $\interpretation$ that, given timepoints $T = t_1, ..., t_{max}$, is any mapping  
$\groundLiteralSet \times T \to \semiLattice$. 
The set~$\interpretationSet$ of all interpretations can be partially ordered via the ordering: $\interpretation_1\preceq \interpretation_2$ iff for all ground literals $\groundedLiteral \in \groundLiteralSet$ and time t, $\interpretation_1(\groundedLiteral, t)\sqsubseteq \interpretation_2(\groundedLiteral, t)$. $\interpretationSet$ forms a complete lattice under the $\preceq$ ordering.
From this, we define a satisfaction relationship ``$\models$'' for temporally annotated facts in the usual manner (i.e., \cite{ks92,ssTAI22,aditya2023pyreason}).

\medskip
\noindent\textbf{Rules and Programs.}  We adopt the temporally extended \emph{GAP rules} from~\cite{aditya2023pyreason}.
If $\nonGroundedformula_0$ is an annotated atom and, $\nonGroundedformula_1,\ldots,\nonGroundedformula_m$ are annotated formulas , then 
\begin{align*}
r\equiv \nonGroundedformula_0 &\xleftarrow[\delay]{} \nonGroundedformula_1\,\wedge\,\ldots\wedge\, \nonGroundedformula_m \tag*{\llap{\text{$\delay \geq 0$}}}
\end{align*}
is called a \emph{GAP rule}.
When a conjunction of annotated formulas in the body is satisfied at time $t$, the annotation of the atom $f_0$ in the head gets updated after a delay, $\delay$.
A GAP rule is called a \emph{fact} when the body is empty, and \emph{ground} when it has no occurrences of variables from $\variableSet$.  
Table~\ref{tab:examplerlsRules} shows examples of rules capturing anomalies that are learned based on observing an agent's routine for two weeks.

\begin{table*}[]

\caption{Rules with intuition for modeling the behavior of humans in a geospatial location}
\label{tab:examplerlsRules}
\begin{tabular}{@{}p{0.68\textwidth}p{0.28\textwidth}}
\toprule
Rule & Natural Language\\

\midrule

$abnormal(A): [0.9,1] \leftarrow_{\delay = 0} education(A):[1,1] \wedge utility(A):[1,1] \wedge \after(utility(A), education(A)): [1,1]$ & If the agent $A$ goes to a $utility$ area like barns and sheds after visiting
an $education$ location (annotated by [1,1])
   like a school, 
   then it shows high abnormal activity by updating its lower bound to $0.9$  \\
\vspace{0.5pt} &\vspace{0.5pt} \\
$abnormal(A): [1,1] \leftarrow_{\delay = 0} industrial(A):[1,1] \wedge assembly(A):[1,1] \wedge \after(assembly(A), industrial(A)): [1,1]$ & It is anomalous that  an agent who works in
   a highly hazardous $industrial$ location will directly go 
   to a highly populated location like theaters ($assembly$) \\
\bottomrule
\end{tabular}
\end{table*}
\medskip

We define a \textit{program}, $\Pi$, as a set of rules and TAFs. An interpretation $I$ satisfies $\Pi$ if for all $e \in \Pi, I \models e$.  We also leverage a fixpoint operator presented in \cite{ssTAI22,aditya2023pyreason} designed for use on lower-lattice annotated logic (which provides analogous results shown earlier for the operator introduced in \cite{ks92}).  
The key here is that the operator provides a minimal model, hence exact answers to entailment under the assumption of consistency.  
As we learn logical rules from data, we can control consistency, so this is a reasonable assumption.  We also use a fixpoint operator as per~\cite{ks92,ssTAI22,aditya2023pyreason} to perform deductive inference.  As per the previous work, fixpoint operator $\fixpointOperator$ is a map from interpretations to interpretations and is applied multiple times until convergence, denoted by $\fixpointOperator^*$.

\medskip
\noindent\textbf{Abductive Inference.}  In this paper, we formalize an abduction problem as: given observations, represented as a set of TAFs denoted $O$, a set of hypotheses, which is also a set of TAFs denoted $H$, program $\Pi$, and parsimony function $\prevalue$ that maps interpretations to positive reals,
the goal is to identify an explanation $E$, a subset of $H$ such that $\Pi \cup E \cup O$ is consistent (in other words, $\program \cup E$ is consistent and entails $O$). Note that as both $E$ and $O$ are TAFs, we are not guaranteed consistency.
Parsimony is measured by $value(\Gamma^\conv_{\Pi\cup E \cup O}(I_\bot))$ (where $I_\bot$ assigns all atoms at all time points to the annotation $[0,1]$, total uncertainty).  An explanation with the lowest value for $value(\Gamma^\conv_{\Pi\cup E \cup O}(I_\bot))$ is an \textit{optimal explanation}.

\begin{example}
\label{ex:graph}

    Building on the notion from example \ref{ex:syn}:
    in our use case, a set of observations $O$ is a set of $TAFs$ (essentially sets constraints on the agent's location at certain times), where each $TAF$ indicates that an agent $a$ is at $loc \in \domainSet_{loc}$ at time $t$.
    $H$ is the set of all possible locations $\domainSet_{loc}$ of the agent. A set of learned temporally extended \emph{GAP rules} $\program$ indicate the normalcy of an agent in a graph $G$. Here, $G$ is a series of $TAFs$ that is formed with nodes from $\domainSet_{loc}$.
    If $\atPred(a,loc): \mu_t$ is a ground atom, then we impose graphical constraints like $\neg \atPred(a,loc'): \mu_{t+1}$, where $\connPred(loc,loc'):\mu_t$ is not true in $G$ and $loc \neq loc'$.
      As seen in Table~\ref{tab:examplerlsRules} we use the lower bound of the annotation in the rule's head to represent the confidence of the body being historically abnormal. When the agent's movements relate to the body, its annotations are updated to $[1,1]$ for time $t$.
    Explanation $E$ is a movement sequence (which is a set of $TAFs$) from $loc_{start}^i$ to $loc_{end}^i$ that fires most rules in $\program$. 

\end{example}

\medskip
\noindent\textbf{Framing the Abduction Problem as a Search Problem.}  
While the general case of such an abduction problem is intractable, we have two key insights that apply to our domain problem.  The first is a structural concern: if we are reasoning about an agent moving in a geospatial setting using a ground vehicle, we know that the agent cannot possibly teleport, so it is restricted to traveling along the graph at a certain speed.  Similar restrictions have been applied to other problems such as social media diffusion~\cite{aditya2023pyreason}, power grid failure modeling~\cite{powergrid}, and knowledge graph completion~\cite{anyBURL2020}.  This allows for TAFs from set $H$ to be selected in a sequential manner while at the same time limiting the TAFs to those consistent with the graphical structure, and the application of search routines such as depth-first search (which we implement in our experiments).  However, we note that this does not reduce the branching factor enough to afford tractability.  In the next section, we present results that allow for a provable lower bound on $value$ in the general case, which we employ in our use case to obtain tractability and further scalability.

\vspace{-1em}
\section{Efficient and Scalable Geospatial Abduction for Trajectory Generation}
\label{sec:theory}

The complexity of logic-based abduction~\cite{Eiter1995complexityLogicBasedAbd,  LIBERATORE201522} can be reduced by using a logic-based parsimony function instead of a standard logic-based function that is intractable due to the number of explanations. We introduce a parsimony function based on the aggregate truth values assigned using a logic program learned from data. The $value$ obtains the lower bound of aggregate over the annotations on an atom $b$ at time $t$ for the minimal model $\interpretation$ of $\program \cup E \cup O$. Using a lower bound on such an aggregate, we obtain an admissible heuristic allowing us to use informed search (i.e., A*). By extension, this addresses both intractability and scalability issues.

\medskip
\noindent\textbf{Bounding the Parsimony Function to enable Informed Search.}  
We now provide new results that allow us to create a lower bound on the parsimony function $value$ by taking a subset of the logic program. These rigorous results imply that we obtain an admissible and consistent heuristic function for A* -- hence, the resulting usage of the lower bound of $value$ in A* can provide an exact solution.  Note that these are general results, not specific to the use case we are studying.  However, clearly, their applicability depends on the subset of $\Pi$ being non-trivial (e.g., the use of $\emptyset$ would be unhelpful).  Further, the idea is that the subset of the logic program also offers a computational advantage. By showing for a given ground atom $b$, time $t$, and $\program' \subseteq \program$ and $I$, $ \Gamma^\conv_{\program'}(I) (b, t) \sqsubseteq  \Gamma^\conv_{\program}(I) (b, t)$, we state the following, which in turn gives us a lower bound on $value$:

\begin{theorem}
\label{thm:adm}
For ground atom $b$, timepoint $t$, $\program' \subseteq \program$, and $I' \preceq I$, 
we have
$\Gamma^\conv_{\program'}(I') (b, t) \sqsubseteq  \Gamma^\conv_{\program}(I) (b, t)$.
\end{theorem}

\medskip
\noindent\textbf{Informed Search Strategy.}   For our use case, a logic program $\program$ is learned where the head of the rules is $\anomPred(agent)$ from Example~\ref{ex:syn}. The body of the rule is determined by two major symbolic landmarks in $G$ that are $n \in \mathbf{Z^+}$ hops away. Consider a single movement as moving and staying in a location (in our case, that is one hop away) from the current location during time $t$. For single hops, a subset of the logic program $\program^{SH} \subseteq \program$ is learned. 
 In the general case, Theorem~\ref{thm:adm} shows that for a subset of the logic program $\program' \subseteq \program$, we get a lower bound on $\prevalue$. For a given set of movements $\interpretation$ in $G$, we employ the $\prevalue$ for a $\program^{SH}$ as the heuristic function in an informed search strategy. We note that the increase in $value$ resulting from single hop rules is inherently modular, meaning that for any node in the frontier set, such quantity is invariant.  This allows us to precompute this increase for all nodes and store it in a graph-based data structure $G_w$.
 Considering the number of iterations of $\Gamma$ as well as grounding for single movements versus a sequence of multiple movements, $\Gamma^\conv_{\program^{SH}}(\interpretation)$ is easier to compute, towards obtaining the heuristic value.

\medskip
\noindent\textbf{Scalable Heuristic Computation.}  Computation of $\prevalue$ can be expensive on a whole logic program $\program$ as it involves computing $\Gamma^\conv_{\program}(\interpretation)$ given a set of movements $\interpretation$ as multiple anomalies can be inferred from multiple sequences of movements. The process of grounding can be expensive computationally, but we can gain efficiency by considering $\program^{SH} \subseteq \program$, where anomalous behavior rises from single movements. Using  Theorem~\ref{thm:adm} we can efficiently prune abnormal candidate movements with informed search.
We precompute the heuristic function by weighting the graph $G$ with the lower bound on $\prevalue$ for all possible single movements and obtain $G_w$. For further scalability, we compute $\prevalue$ in a need-based fashion called {\em ad-hoc weighting} instead of precomputing it for all possible single movements in $G$. During heuristic computation we obtain the lower bound of $\prevalue$ only when a certain movement is needed. For ad-hoc weighting, we compute $\prevalue$ considering the agent's frontier up to a depth of $1$ at each step. 

\vspace{-1em}
\section{Software Stack}
\label{sec:software}

\begin{figure}[t]
\centering
  \includegraphics[scale=.28]{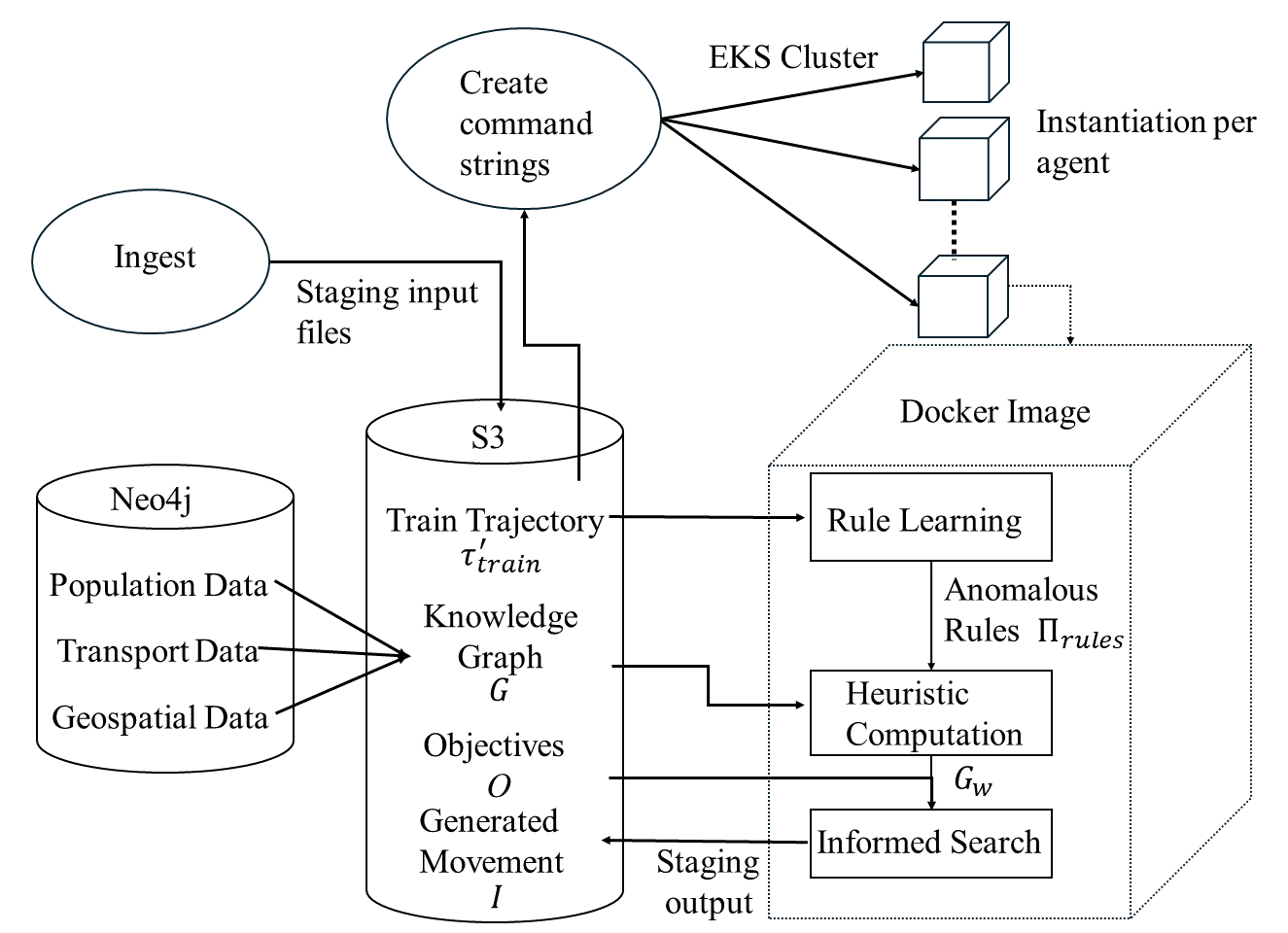}
  \caption{Visual representation of deployment.} 
  \label{fig:dag}
\end{figure}

Our logic program-guided abduction strategy is part of a software stack that was deployed in a cloud environment on Amazon Web Services (AWS) as per mandate by the government customer.  Figure~\ref{fig:dag} depicts the overall system architecture
(DAG structure).

\medskip
\noindent\textbf{Overall Workflow.} 
The pipeline interfaces with the government system to access the raw geospatial data with related knowledge for 4 geolocations as well as training agent trajectories (as seen in Figure \ref{fig:data} (left)) and required objectives for each agent (framed as a set of observations $O$). The required objectives include typical human activities of single (visiting a friend, restaurant, etc) and recurring (going to work, etc.) movement types. 
Initially, we host the raw geospatial data in a Neo4j server and form a consolidated knowledge graph $G$ (symbolic landmarks extracted from $G$ are seen by the symbols in Figure~\ref{fig:data} (left)). This is stored in an S3 bucket also containing the training data (consisting of both trajectories and objectives for each agent). For each agent, we extract spatial and temporal constraints from the training data. Based on a single trajectory of an agent, we learn a set of anomalous rules $\Pi_{rules}$ (as seen in Table~\ref{tab:examplerlsRules}). Both $G$ and $\Pi_{rules}$ are used to compute the heuristic values by creating a weighted graph $G_w$.  
We also use constraints to perform an informed search algorithm to generate a normal trajectory (as seen in Figure~\ref{fig:data} (right)). 

\medskip
\noindent\textbf{Data Ingest.}  Our initial ingest and staging containerized processes are held in the DAG as nodes. Our ingest mechanism first parses the objective files for the agent to determine the locations of corresponding training trajectories. Secondly, based on the geolocation, we retrieve the appropriate knowledge graph and link it to each agent. Finally, we process the data associated with each agent (objectives, training data, graph) to a predefined staging area in the S3 bucket.

\medskip
\noindent\textbf{Instantiation.} This step analyzes the staging folders and creates the necessary string commands specific to each agent. We launch pods for all agents with a movement-generation Docker image to process its particular objectives. As the container runs, generated movement instruction files are pushed to the appropriate output directory.

\medskip
\noindent\textbf{Rule Learning.} To generate normal movement, we learn rules from the agent's historical data capturing realistic behavior. Sequences of movements deviating from this behavior are considered anomalous. Inferring from longer sequences can make the algorithm computationally expensive, but when we only compute a subset of rules involving shorter sequences, we can efficiently prune candidate movements that are highly abnormal using search algorithms.  There are different kinds of rule types, which we call single-hop and multi-hop rules ($\program_{SH}, \program_{MH}$). Here, $\program_{SH}$ leverages single movement frequency (for instance Table~\ref{tab:examplerlsRules}) while $\program_{MH}$ leverages a set of multiple movements (in a similar format of Table~\ref{tab:examplerlsRules}). The logic program $\program$ includes both types of rules. From Theorem~\ref{thm:adm}, we can get
$value(\Gamma^\conv_{\Pi^{SH}\cup E \cup O}(I)(b, t)) \sqsubseteq value(\Gamma^\conv_{\Pi\cup E \cup O}(I)(b, t))$ as $\program^{SH} \subseteq \program$; this is demonstrated in Figure \ref{fig:baseline}.

\medskip
\noindent\textbf{Search.} We use PyReason~\cite{aditya2023pyreason} to compute the fixpoint operator $\Gamma$ used for both the actual calculation of value and the creation of the heuristic function. We use the aggregate function as an intersection over all annotations of the rules fired by $\Gamma^*$.
This is computed in an ad-hoc fashion for possible movements, and we weight the graph based on those movements to form $G_w$. Given the set of constraints $O$, we form sub-abduction problems to perform an A* search, and generate normal trajectories satisfying all required objectives.

\vspace{-1em}
\section{Internal Evaluation}
\label{sec:internaleval}

\noindent\textbf{Experimental Setup.}  
The government provided us with simulated data we used for the internal development of our approach. We leverage three main kinds of data. Firstly, curated data for four geospatial locations: Knoxville, Singapore, Los Angeles, and San Francisco is collected from multiple source datasets, namely USAStructures~\cite{usastructure}, Planetsense~\cite{planetsense} (which provides geospatial information), Open street map road network (which gives transportation information), Urbanpop~\cite{urbanpop} (containing population data), and other data collected as part of the program. Secondly, we have simulated trajectory data of 40,000 human agents across all 4 locations, which mimic realistic human activity. Four different teams each provide a different simulation environment for generating realistic training data. Note that this data comprises only location data and does not include information on actual people. Thirdly, we have a set of objectives for each agent that specifies spatio-temporal constraints.
From the curated data, we use three types of input data: geospatial (building types, occupancy, etc), population (census data, population survey data, etc), and transportation (road network, statistics on traffic flow, etc) to build a consolidated knowledge graph $G$. Each node in $G$ is either an intersection point from the road network or has attributes that convey its landmark category resembling building occupancy such as Commercial (stores, parking), Unclassified (does not require much security, non-residential), Non-Profit (general offices, Emergency Operation Centers), Residential (apartments, hotels), Assembly (convention centers, stadium), Education (libraries, schools), Utility (barns, water treatment), Industrial (hazardous factories- chemical factories, metal processing factories, construction), Agriculture (agricultural use land), Government (military, fire station).  
Each trajectory is a sequentially ordered tuple of length 2 weeks, consisting of latitude, longitude, and timestamp indicating the agent's location at each time point (cf.\ Figure~\ref{fig:data} for an example of the trajectory on the graph). Moreover, every set of objectives describes spatial and temporal constraints on the trajectory to be generated. We generated 38 trajectories that satisfy all constraints for which the objectives were provided.

We conducted all experiments on a high-memory CPU machine with 128 cores, and 2000GB memory, using PyReason software \cite{aditya2023pyreason} for inference. Rules (similar to Table~\ref{tab:examplerlsRules}) were learned using a bottom-up technique comparable to related work~\cite{APTL} where you restrict the body of the rule to contain historically possible single movements.

\begin{figure}[t]
\centering
    \includegraphics[scale=.43]{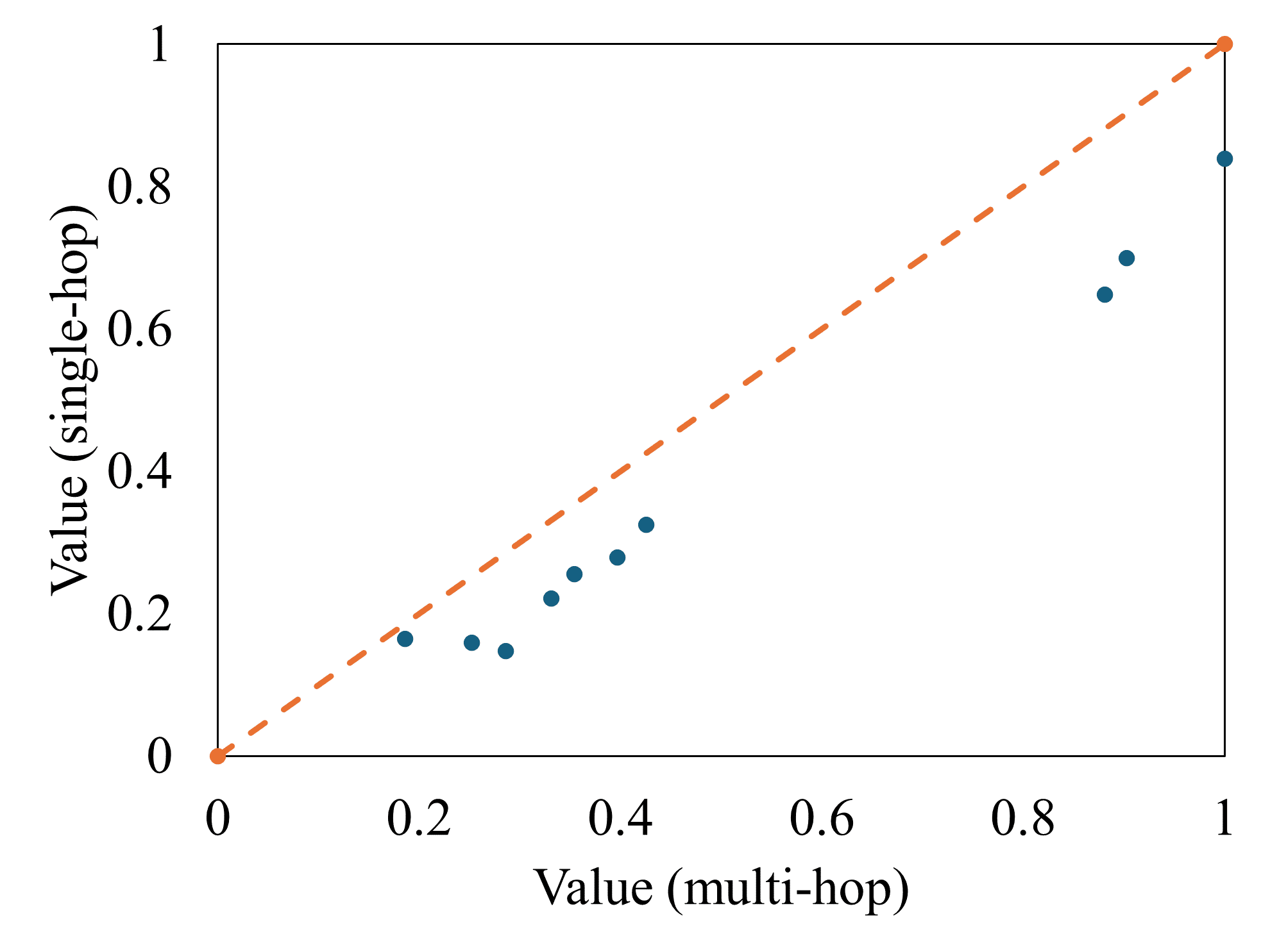}
  \includegraphics[scale=.14]{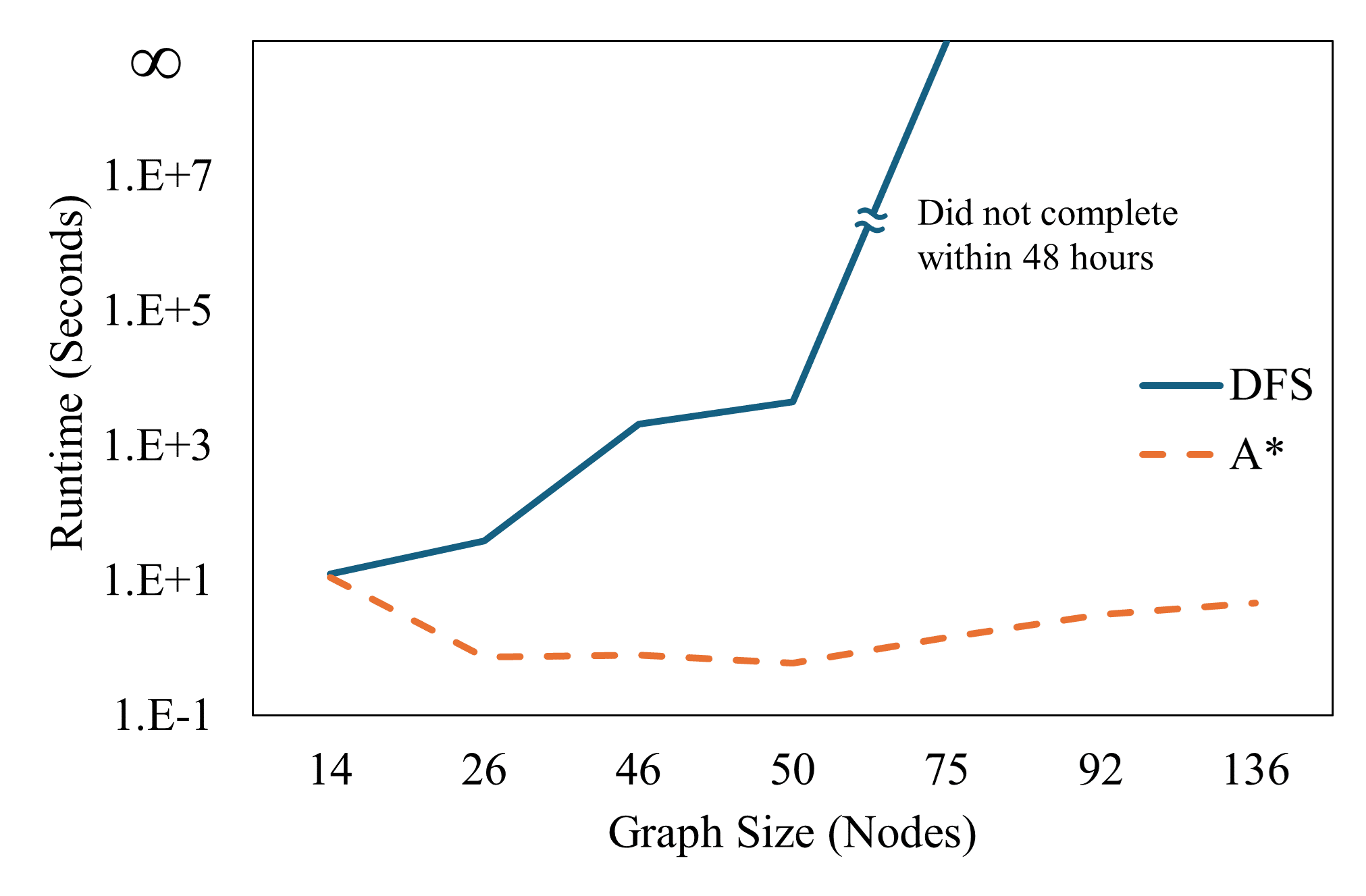}

    \caption{Left: Value with subset logic program, logic program $\program', \program$ for the Knoxville location.  Right: Runtime Comparison with Depth-First Search and A* Search.}
    \label{fig:baseline}
\end{figure}

\medskip
\noindent\textbf{Empirical Validation of Theoretical Claims.}  We present the results of two experiments as seen in Figure~\ref{fig:baseline} to validate our claims concerning the use of the heuristic function and its employment as part of an informed search strategy. 
Figure~\ref{fig:baseline} (left) shows that the heuristic value computed by the subset of the logic program is lower than the actual value in all our experiments since the data points lie below the dashed line -- this aligns with Theorem~\ref{thm:adm}. On the right panel, we see that the heuristic effectively addresses the tractability issues of the search, where the search conducted with the depth-first search algorithm without the heuristic could not be completed in under 48 hours even on small graphs with 50 nodes. Our approach using a heuristic maintains a lower running time as shown in Figure~\ref{fig:baseline} (right).

\begin{figure}[t]
\centering
   \includegraphics[scale=.16]{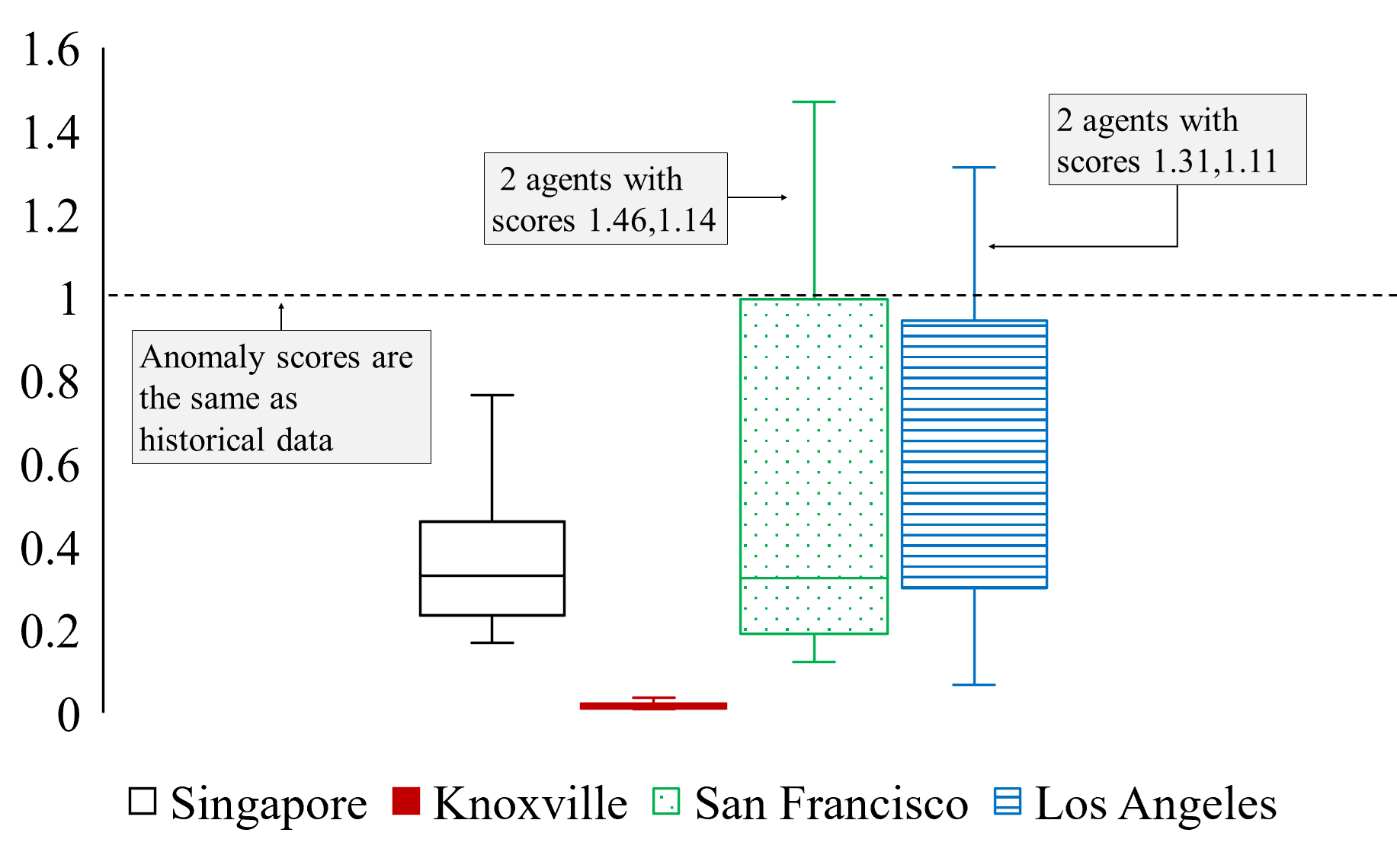}
  \includegraphics[scale=.45]{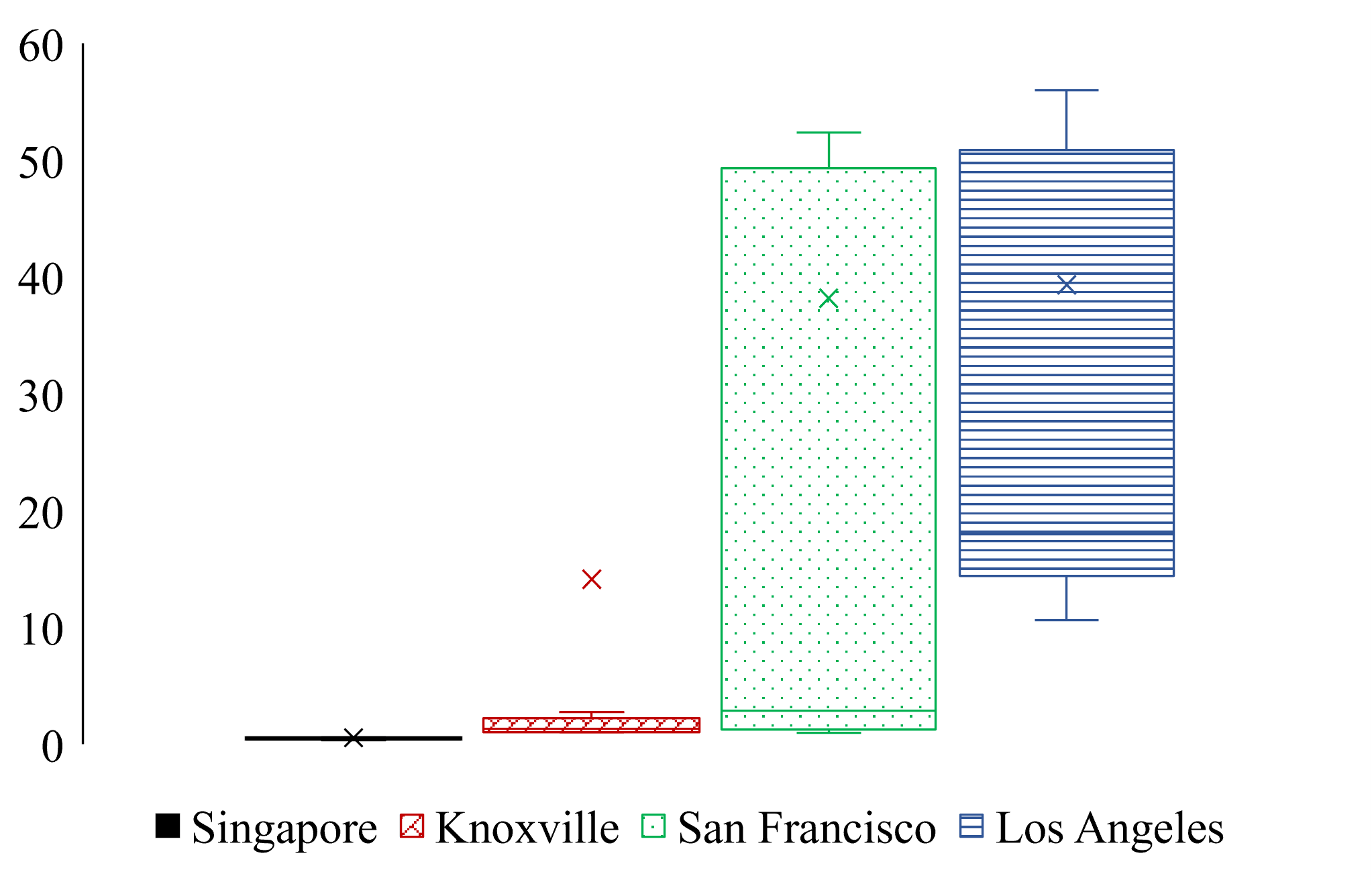}

    \caption{Left: Relative Anomaly Ratio. Anomalies in generated movements when compared with anomalies found in the training data by an ensemble ML anomaly detection algorithm. Right: Runtime speedup due to ad-hoc weighting.}
    \label{fig:rarruntime}
\end{figure}

\medskip
\noindent\textbf{Trajectory Robustness.}  A key aspect we wish to achieve with the generation of realistic movement trajectories is their robustness to anomaly detection. 
In this experiment, we run an ensemble of machine learning methods to perform anomaly detection over the generated and training trajectories to get their anomaly scores.  The anomaly score of the ensemble is then compared with the average of that score over the historical trajectories of the agent, resulting in the relative anomaly ratio. In Figure \ref{fig:rarruntime} (left), we can see a box plot of anomaly scores of generated trajectories relative to the training data. In $90\%$ of the data, a ratio lower than $1$ is observed, indicating that mostly, anomalous movements identified in generated trajectories are no more frequent than those occurring in the training trajectories.
For the cases where the anomaly detector found more anomalies than in the training data, the proportions are $46\%, 31\%, 14\%$, and $11\%$, which is likely to be lower than a practically employed high-precision threshold.

\medskip
\noindent\textbf{Scalability.} 
We implemented both ad-hoc and non ad-hoc weighting for all 4 locations -- see results in Figure~\ref{fig:rarruntime} (right). 
Ad-hoc technique gave a maximum speedup of 245.42 and a minimum of 0.32 when compared to the non ad-hoc technique. Over all, ad-hoc weighting is beneficial due to reduced runtimes. Several outliers (not depicted in the plot) for 3 AOIs were observed with speedups of $128.05, 245.04,$ and $140.46$.
Furthermore, when the graph size (the number of edges) increased by a factor of $4.69$, the speedup was boosted by a factor of $72.05$. The median speedups as the graph gets denser are $0.44,1.30, 2.86$, and $17.98$. These improvements show that the ad-hoc technique will scale well as the graph size grows significantly or gets denser when compared to non ad-hoc technique. 
In many cases, the non ad-hoc technique did not complete after running for more than 10 days, suggesting a speedup of over $1,000$x for those cases. Though such results are favorable to our approach, we did not consider these samples in the determination of the numbers in this section.

\begin{figure}[t]
\centering
  \includegraphics[scale=.07]{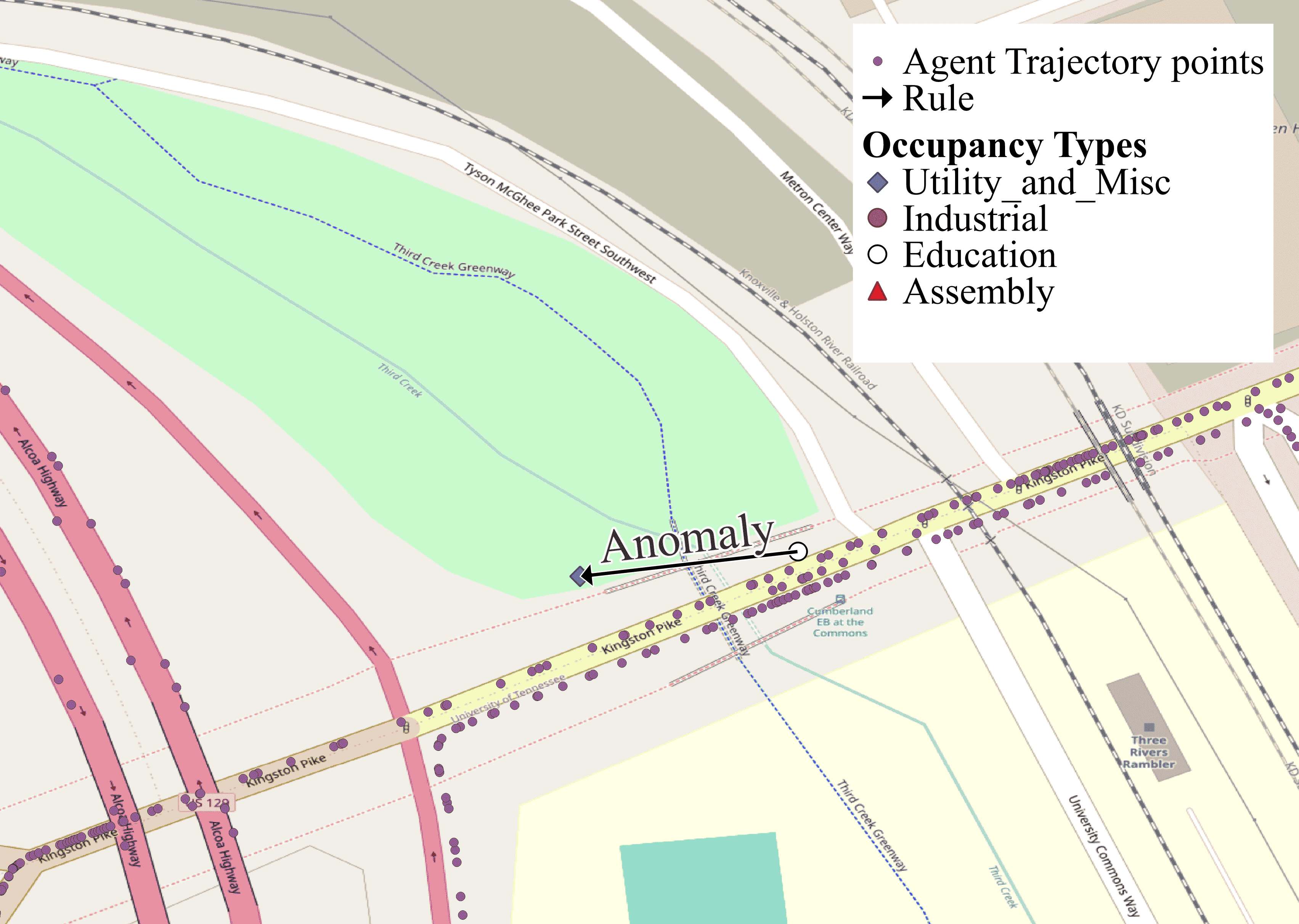}
  \includegraphics[scale=.053]{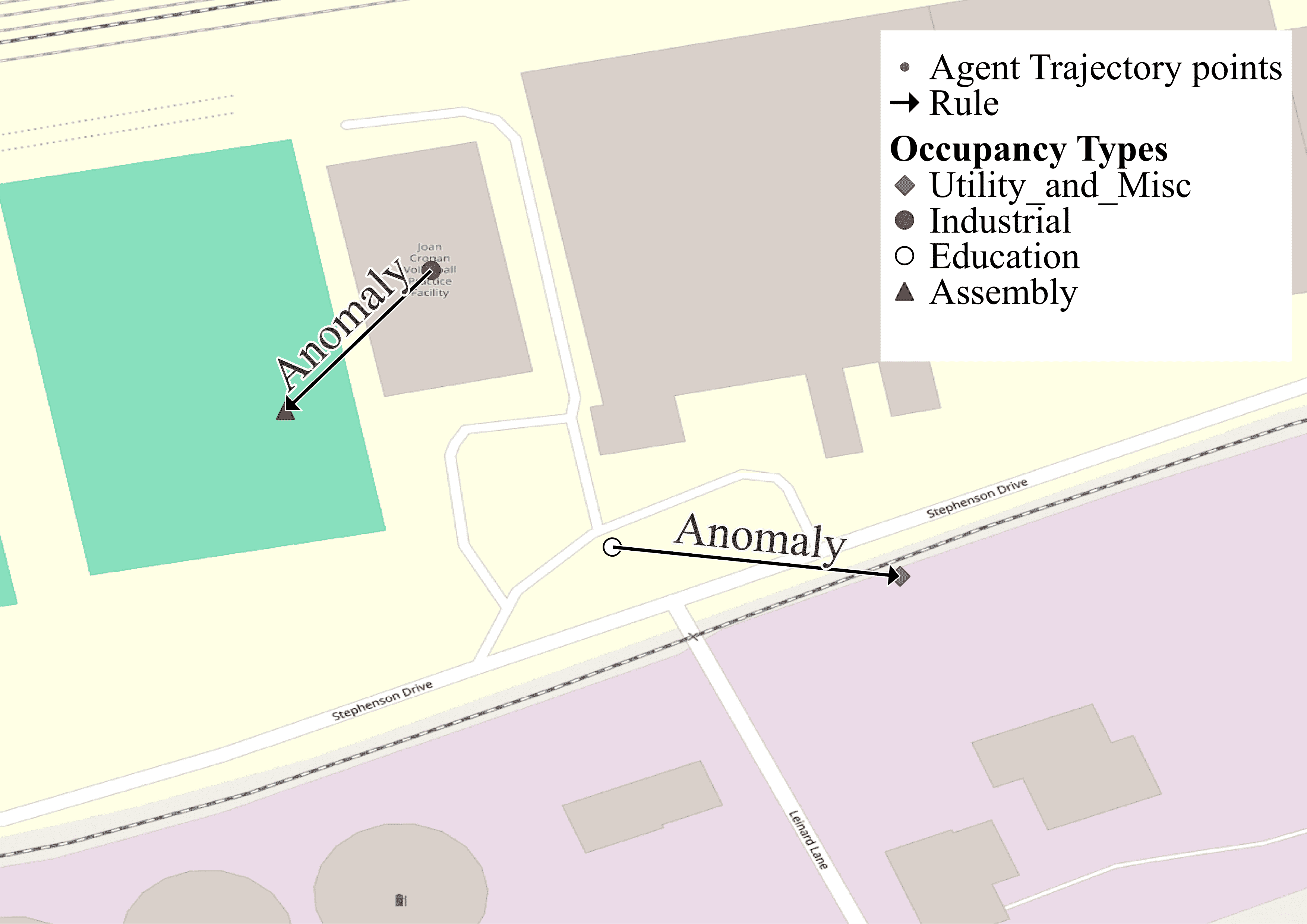}
    \caption{Visual depiction of rules from Table~\ref{tab:examplerlsRules}. Left: The top rule. Right: Both rules. }
    \label{fig:xrule}
\end{figure}

\medskip
\noindent\textbf{Explainability.}
Movemments in a trajectory are generated using a fixpoint-based algorithm~\cite{ks92} on $\program$. Movements tagged anomalous by a detected in such a trajectory can be backtracked to specific rules. Due to the modularity inherent to the approach, these rules assist analysts in assessing particular anomalous behaviors.
An analyst can explore the anomalies detected on the generated trajectory by tracing back to the rules that caused such movements; for example, cf.\ Figure~\ref{fig:xrule}. 
Note that trajectory generation on historical data with our approach does not let an external entity infer an agent's identity. Even though the program is learned from the data, trajectory points in the agent's generated trajectory needn't be the same as historical data, which also contributes to privacy protection.

\vspace{-1em}
\section{Deployment for Independent Evaluation}
\label{sec:externaleval}

\noindent\textbf{Independent Evaluation Setup.}
The generated trajectories are evaluated against nine different anomaly detection algorithms, each created by a government contractor. Further, these evaluations take place in the 4 locations examined in Section~\ref{sec:internaleval} but with newly simulated historical data for a duration of 2 months not previously available to our team. The results were collected and assessed by the government test and evaluation team using government-created cloud infrastructure to receive our outputs.
The evaluation teams included an external T\&E team, four simulation teams that run simulation engines to provide training data replicating real-world dynamics innate in human movement, and other teams that designed bespoke anomaly detection algorithms to evaluate the generated trajectories. Additionally, with the same simulators more movements are synthesized (which we are blind to) to be inserted with the 113 generated trajectories on the newly simulated historical data.

\begin{figure}[t]
\centering
  \includegraphics[scale=.35]{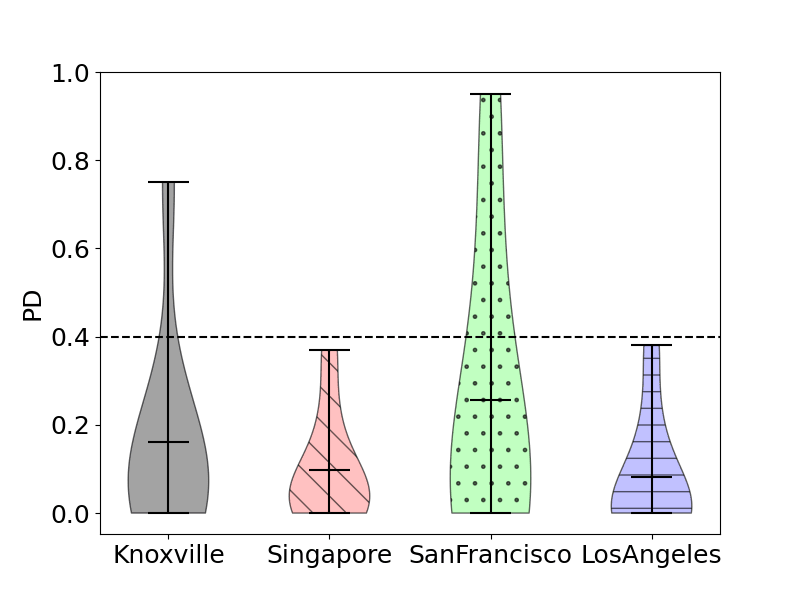}
  \includegraphics[scale=.35]{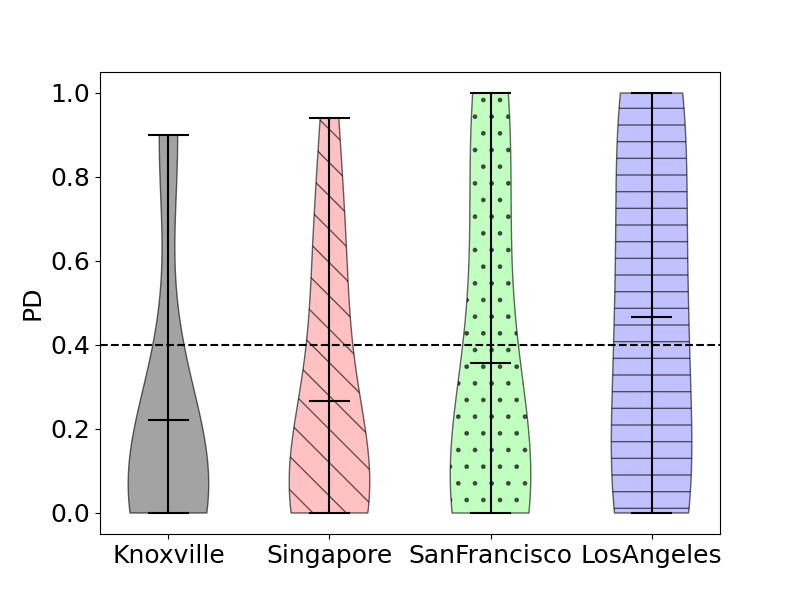}
    \caption{Prob.\ of detection (PD) from independent trials against ML-based anomaly detectors.  Left: PD based on agents discovered.  Right: PD based on points discovered.}
    \label{fig:deploy}
\end{figure}

During this trial, we were evaluated on the probability of detection, defined as per the government's guidance~\cite{baa}: $PD = \text{Correctly Detected Anomalies}/ \text{All Anomalies}$.
In Figure~\ref{fig:deploy} we show the PD of generated movement trajectories against the various machine learning anomaly detectors developed in the program and for each of the four simulation environments.  We note that several anomaly detectors were developed by teams that created simulation environments, which appeared to have an advantage  
We present results for F1-optimized anomaly detectors in Figure~\ref{fig:deploy}. It
 shows results curated by the government T\&E team where anomalies are defined as anomalous agents being identified as well as when defined as points in the agent's trajectory. Note that our approach was designed to identify major intersections or landmarks that the agent would traverse.  An interpolation procedure - not part of the overall route planning was used to fill in waypoints between such major intersections.  This is likely why many anomaly detectors were able to identify anomalous points (Figure~\ref{fig:deploy} right) but generally failed to find the agents from the overall trajectory (Figure~\ref{fig:deploy} left).  Improvement to the interpolation procedure is an area of future work.  In the majority of cases, we met the government standard of having a PD below $0.40$.

\vspace{-1em}
\section{Related Work}
\label{sec:relwork}

\noindent\textbf{Relation to Abduction Literature.}  Traditionally, in an abductive inference problem, an \textit{explanation} is determined based on a model, which imposes some form of constraints, and a set of observations that though observed are phenomena that would be inferred from the model and explanation. The hardness of logic-based abduction (e.g., when the model is some form of logic program) is well established~\cite{Eiter1995complexityLogicBasedAbd,LIBERATORE201522}, and a key source of this complexity is that the number of explanations is intractable in the general case.  As a result, \textit{parsimony} requirements are used to specify criteria over which explanation is preferable~\cite{peng-reggia90} -- for example, previous work on geospatial abduction (but not abducing trajectories) minimized cardinality as a parsimony requirement~\cite{shak11}.  In another example, recent applications of abduction to machine learning, such parsimony requirements are expressed in terms of training loss~\cite{huang2023enabling,dai2019bridging}.  Our work differs from all of the aforementioned papers in that we introduce a parsimony requirement based on the aggregate truth values assigned by a logic program that can be learned from data.

\medskip
\noindent\textbf{Relation to Literature on Geospatial Trajectory Generation.}  
The abundance of trajectory data permits the analysis of mobility patterns and trajectories such as Markov-based paradigms have been applied for momentary goals (location prediction) \cite{markov2012, lu2013approaching} and coarse-grain human movement analysis~\cite{wang2024cola, doi:10.1073/pnas.2313171120, alis2021generalized} - both different problems than this paper.  Recent studies used recurrent neural networks (RNN)~\cite{rnn}, generative adversarial networks (GAN)~\cite{gan,gan2}, graph neural networks (GNN)~\cite{gnn}, and transformer-based approaches~\cite{wang2024cola} on real datasets to generate trajectories. Though these approaches tend to capture spatial-temporal correlations, they lack explainability.  Recurrent-based approaches generate trajectories with short duration and spatial coverage.
Adversarial approaches need more training data to get realistic outcomes while GNN-based models are time-consuming.  
A reinforcement-learning approach for sequence modeling~\cite{janner2021offline,chen2021decision} can be extended for human trajectory generation but has challenges for large graphs. They heavily rely on the structure of the reward function. 
Interpretability has been induced using the latent space to see the distribution of uncertainty of semantic concepts~\cite{itkina2022interpretable}.  
Our approach can generate long-range trajectories spanning over a city with a length of two months, in a data-efficient manner~\footnote{We had a single 2-week training sample per agent.  For comparison, the ML approach of \cite{ gan}, uses on average $60$ examples per agent that were 3 years in duration.} and scales to denser graphs. The modularity of our approach allows for detailed explanations.

\vspace{-1em}
\section{Conclusion}
\label{sec:conclusion}

In this paper, we described a system that generates realistic but synthetic human movement trajectories using abduction guided by a logic program.  This system was recently deployed for independent government testing.  We note that in our current iteration, there were several limitations.  For example, we did not employ ad-hoc graph weighting (essentially a sequential operation) and parallelization together -- finding how to achieve the right balance between the two techniques can result in further scalability.  We also look to extend our logical language to provide additional insights into anomalies; for example, time of day is not currently considered, so an anomaly detector explicitly considering that aspect will readily find our trajectories.  Further, we also look to leverage the ability of our underlying logic to accept arbitrary functional symbols to assign truth, allowing to leverage neurosymbolic techniques~\cite{shakarian2023neuro} to directly integrate ML anomaly detectors.

\medskip
\noindent\textbf{Ethics Statement.} 
This work is part of the IARPA HAYSTAC program, which is designed to create simulated environments with generated human movement patterns not associated with actual persons and enable further study of human movement trajectories without relying on actual human data.

\vspace{-1em}
\medskip
\paragraph{\bf Acknowledgement} This research is supported by the
Intelligence Advanced Research Projects Activity (IARPA) via the Department of Interior/Interior Business
Center (DOI/IBC) contract number 140D0423C0032. The U.S. Government is authorized to reproduce
and distribute reprints for Governmental purposes notwithstanding any copyright annotation thereon.
Disclaimer: The views and conclusions contained herein are those of the authors and should not be
interpreted as necessarily representing the official policies or endorsements, either expressed or implied,
of IARPA, DOI/IBC, or the U.S. Government. 
Also, this work was funded by ONR grant N00014-23-1-2580.

\nocite{*}
\bibliographystyle{eptcs}
\bibliography{generic.bib}
\end{document}